\documentclass{article}

\usepackage{graphicx}
\usepackage{array}

\begin{document}


\title{{\bf$^{136}Sn$ and three body forces}}

\author{{\bf M. Saha Sarkar \footnote{maitrayee.sahasarkar@saha.ac.in}, S. Sarkar  \footnote{ss@physics.becs.ac.in}} \\
{\it Saha  Institute  of  Nuclear  Physics, Bidhannagar,
Kolkata-700064, INDIA}\\
{\it Indian   Institute   of   Engineering  Science  and
Technology, Shibpur, Howrah - 711103, INDIA}}

\maketitle

\begin{abstract}
New   experimental  data  on  2$^+$  energies  of  $^{136,138}Sn$
confirms  the  trend  of  lower  2$^+$  excitation  energies   of
even–-even  tin  isotopes  with  N  $>$ 82 compared to those with
N$<82$. However, none of the theoretical predictions  using  both
realistic  and  empirical interactions can reproduce experimental
data  on  excitation  energies  as   well   as   the   transition
probabilities  ($B(E2;  6^+  \rightarrow  4^+$)) of these nuclei,
simultaneously, apart from one whose matrix  elements  have  been
changed   empirically   to  produce  mixed  seniority  states  by
weakening pairing. We have shown  that  the  experimental  result
also  shows  good  agreement  with the theory in which three body
forces have been included in a  realistic  interaction.  The  new
theoretical   results   on  transition  probabilities  have  been
discussed to identify  the  experimental  quantities  which  will
clearly distinguish between different views.

\end{abstract}




\section{Introduction}

Nuclei  around  doubly  closed  $^{132}Sn$ lie on or close to the
path of astrophysical r-process flow. Structure of these  nuclei,
particularly  the  binding  energy (BE), low-lying excited states
and  beta  decay  rates  at  finite  temperatures  are  important
ingredients   for  nucleosynthesis  calculation.  However,  these
nuclei  are  usually  experimentally   inaccessible   by   common
techniques   of   nuclear   spectroscopy.   So  far  experimental
investigations have  been  performed  using  spontaneous  fission
sources,  thermal  -  neutron  -  induced  fission deep inelastic
reactions,  fragmentation  and  fission   at   intermediate   and
relativistic  energies. Reactions with neutron – rich radioactive
ion beams are expected to generate very important, reliable  data
base for these nuclei.

So  far  the  experimental  status  \cite{nndc}  was  not  at all
satisfactory for Sn  isotopes  beyond  $^{132}Sn$.  Spectroscopic
information,  such  as  BE  and  low  lying  spectrum,  is  known
experimentally only for $^{134}Sn$. Half-lives of  $^{135-137}Sn$
have  been measured through $\beta$-n decay process. Lifetimes of
these  nuclei  are  very small and production rates are also very
low presenting  challenges  to  spectroscopic  studies.  Reliable
theoretical   results   are   therefore   necessary  and  useful,
especially as inputs to calculations for astrophysical processes.
Being close to the doubly closed neutron-rich $^{132}Sn$ nucleus,
this region therefore has been taken  as  a  fertile  ground  for
testing  nuclear  shell  model  (SM) with interactions suited for
nuclei far from stability.

\begin{figure}
\includegraphics[width=\textwidth,angle=0]{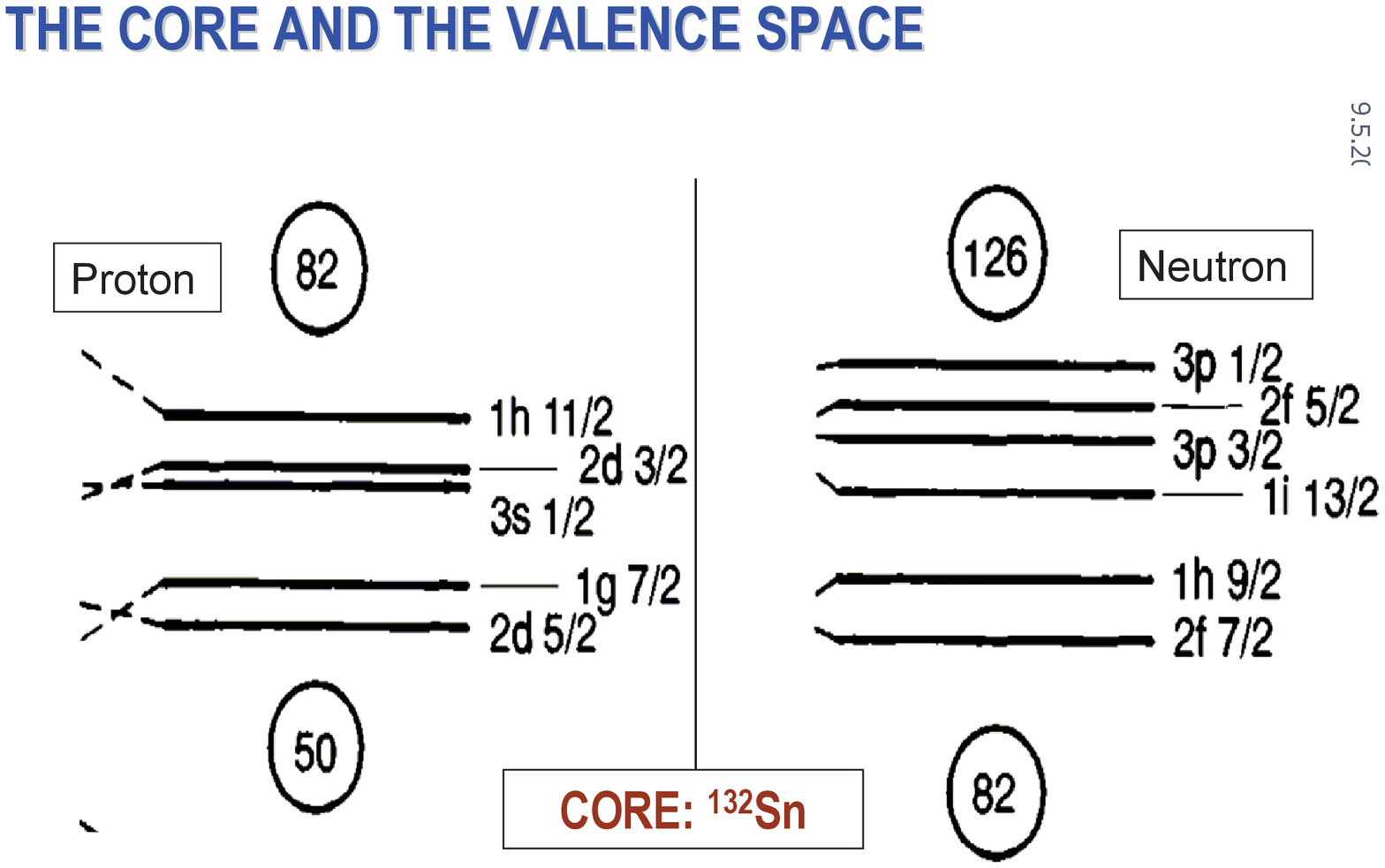}
  \caption{  The  valence  space  for proton (p) and neutrons (n)
above $^{132}Sn$ core.}
\label{basis}
 \end{figure}

\section{Evolution of our endeavours}

\begin{figure}
\vspace{4cm}

\includegraphics[width=\textwidth,angle=0]{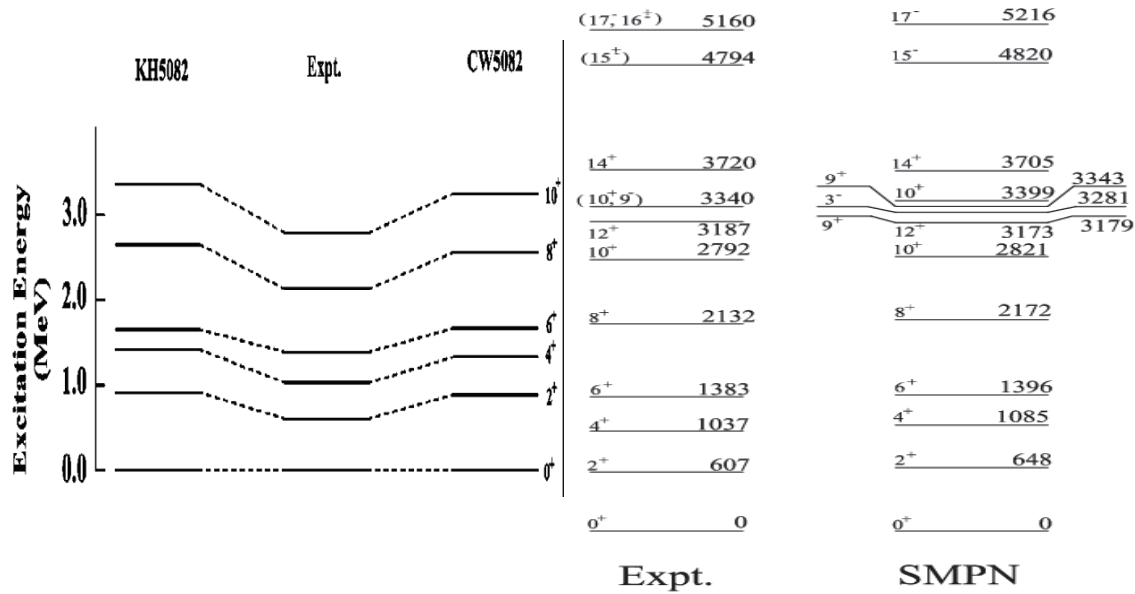}
\vspace{-4cm}
  \caption{  Comparison  of  experimental  and theoretical energy
spectra of $^{136}Te$. }
\label{136te}
\end{figure}

We have started working in this mass region in early 2000, with a
systematic  study  \cite{smp2001}  of  this mass region utilising
available interactions. We have used then  available  version  of
the  shell  model  code  OXBASH  \cite{oxb:1}.  The  interactions
\cite{chou}  used  by  us  were  KH5082  and   CW5082.   The   SM
calculations  were  carried  out  in the Z50N82 model space (Fig.
\ref{basis}). The earlier workers realised that  there  are  many
points  of  similarity  between  the  spectroscopy  of the doubly
closed shell regions around stable  $^{208}Pb$  and  neutron-rich
$^{132}Sn$.  The two interactions mentioned above are such (1+2)-
body  nuclear  interaction  Hamiltonians  where  interactions  of
$^{208}Pb$  region  are  scaled  down to the next lower available
doubly magic domain, i.e., in the $^{132}Sn$ region. Most of  our
results  satisfactorily explained the available experimental data
for  several  neutron  rich  isotopes  of  $_{52}Te$,   $_{53}I$,
$_{54}Xe$,  $_{55}Cs$,  $_{56}Ba$  with  N=82 and 83. However, we
found for  N$\geq$84  isotones,  the  interactions  were  not  as
successful and seemed to be inappropriate. The evergy levels were
overpredicted  for  these neutron rich nuclei (Fig. \ref{136te}).
So  the  limitations  of  the  two  interactions  in   predicting
experimental observables in N$\geq$84 isotones, clearly indicated
a  necessity of changing particularly the $n - n$ two body matrix
elements (tbmes) of the interactions.

\subsection{Construction of the new Hamiltonian}
Keeping  the  conclusion of our first work in this mass region in
mind,  we  attempted  a  simple  modification   of   the   CW5082
interaction \cite{chou}. The CW5082 Hamiltonian was
modified  in  the  light  of  available  information  on  binding
energies, low-lying spectra of two-nucleon nuclei: A=134  isobars
of  Sn,Sb  and  Te  isotopes. The energies of the single particle
orbitals of the valence space above the $^{132}Sn$ core have been
replaced  by  the  experimental  ones.  The   details   of   this
modification    procedure    have    been   discussed   in   Ref.
\cite{smp2004}. The new  Hamiltonians  work  remarkably  well  in
predicting    binding    energies,    low-lying    spectra    and
electromagnetic transition probabilities for N=82,83 and even for
N $\geq$ 84 isotones of Sn,Sb,Te,I,Xe and Cs nuclei. Out of total
two thousand one hundred and one (2101) tbmes,  only  twenty  six
(26)  tbmes are modified in SMPN interaction. Twelve (12) of them
are $n-p$ tbmes, four (4) are $p-p$ and the rest are ten (10) are
$n-n$ tbmes. Six of these  neutron-neutron  matrix  elements  are
diagonal  elements  coupled  to $0^+$ spin. These matrix elements
are  reduced,  implying  a  reduction  of  the  pairing  this  in
neutron-rich  domain.  The  modified  two body matrix elements of
SMPN are tabulated in Ref.\cite{daess}.

\subsection{Comparison with realistic interaction}
The  success  of  his new interaction prompted us to apply it for
several    new    data    sets   obtained   by   experimentalists
\cite{136I,138I,138Cs}. In the mean time, few other  groups  have
worked   with   CD-Bonn  potential  based  realistic  interctions
\cite{brown, covello, kartamyshev}.  Their  results  also  agreed
well  with  many  of  the  available data. So our next aim was to
compare  the  results  obtained  with  the  empirically  modified
inteactions with those obtained with the realistic interactions.

We   have   used   the  CD-Bonn  potential  based  realistic  CWG
\cite{brown}  and  the   empirical   SMPN   interactions.   These
Hamiltonians  are  used  in this neutron-rich region and have the
same set of single-particle energies of the valence orbitals  but
different  sets  of  two-body interaction matrix elements. The SM
calculations \cite{smp2008} were carried out in the Z50N82  model
space (Fig. \ref{basis}) with the SMPN and CWG Hamiltonians using
the code OXBASH and NuShell \cite{oxb:1}.

\begin{figure}
\includegraphics[width=\textwidth,angle=0]{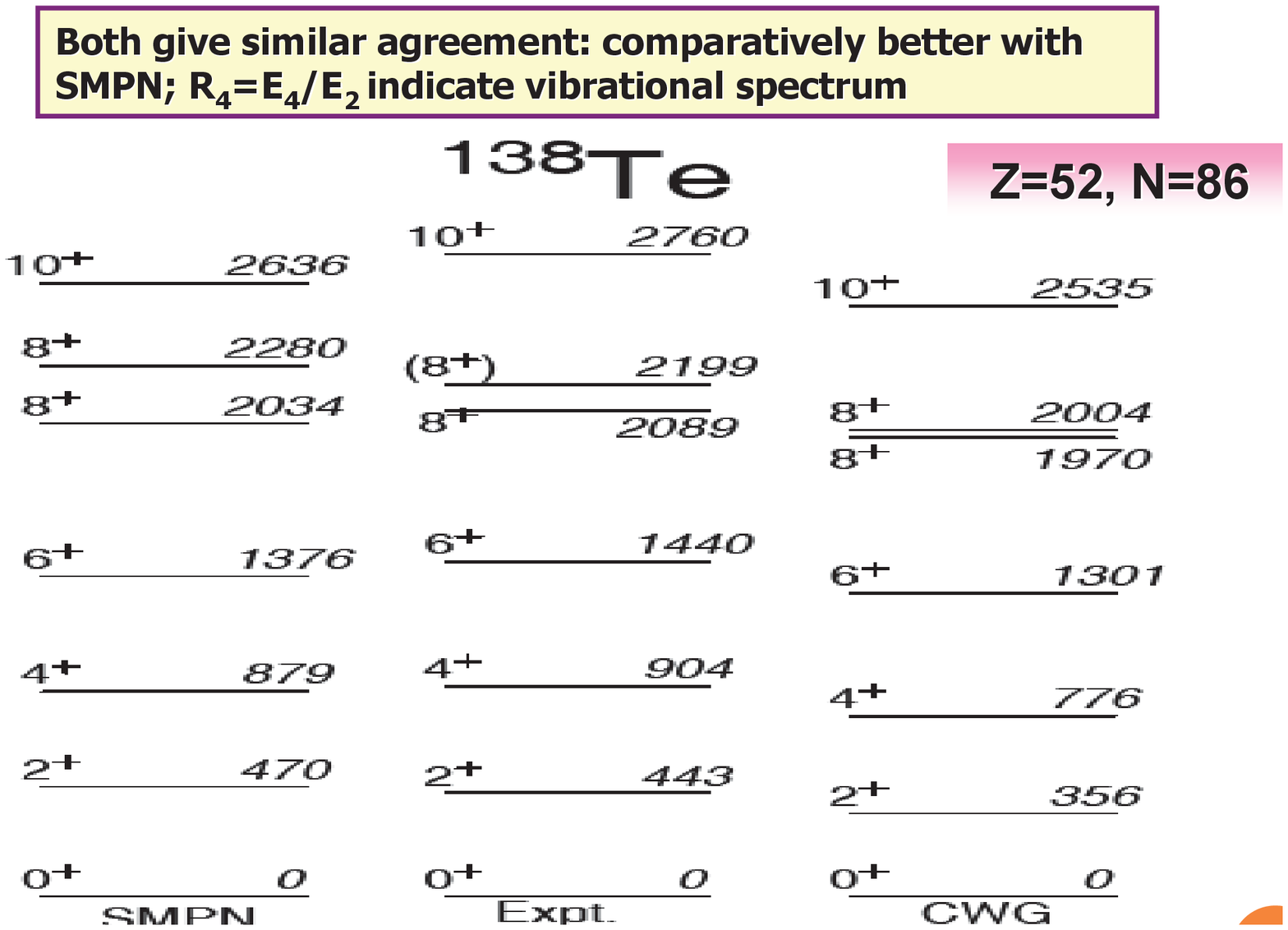}
  \caption{  Comparison  of  experimental data and theoretical
predictions for excitation energies in $^{138}Te$.}
\label{138te}
\end{figure}

 The  calculations  have  been done \cite{smp2008} for $^{138}Sn$
($T_z$=+3.0) with 6  neutrons,  $^{138}Te$  ($T_z$=+1.0)  with  4
neutrons  and  2 protons, $^{138}Xe$ ($T_z$=-1.0) with 2 neutrons
and 4 protons and finally, $^{138}Ba$ ($T_z$=-3.0) with 6 valence
protons. Untruncated  shell  model  calculations  for  6  valence
nucleons  in  this  basis space usually involve matrices of large
dimensions. Calculated energy eigenvalues of the levels with  the
SMPN and CWG Hamiltonians are in good agreement with experimental
ones  up  to  the  highest  observed  level  in  $^{138}Te$ (Fig.
\ref{138te}), $^{138}Xe$ and $^{138}Ba$ \cite{smp2008} (including
the non-yrast levels in $^{138}Xe, Ba$). In the CWG  results  all
the  energy  eigenvalues of the excited states are systematically
underpredicted by about 100-150 keV. For neutron rich  ones  like
$^{138}Te$,  both  these calculations reproduced quite reasonably
well the vibrational spectrum of this nuclei  (Fig.  \ref{138te})
indicated  by its $R_4$= E($4^+_1$)/E($2^+_1$) $=$ 2.04 value. In
both the results, the lowest proton and neutron orbitals have the
maximum occupations. The  CWG  wavefunctions  have  comparatively
larger  occupations  in  other  orbitals  also, indicating larger
configuration mixing.

\subsection{Predictions for neutron rich heavy Sn isotopes}

\begin{figure}
\hspace{-1cm}
\includegraphics[width=.7\textwidth,height=0.7\textheight,angle=-90]{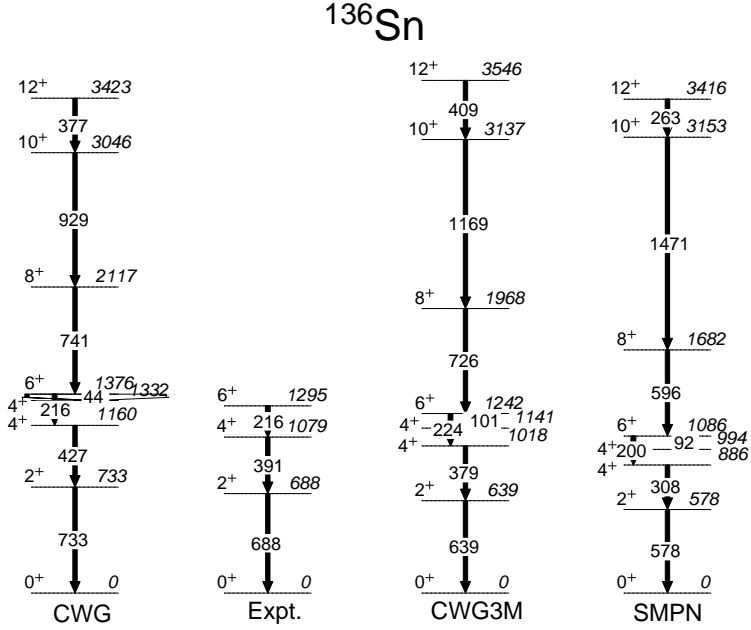}
\caption{  The  calculated excitation spectra of $^{136}Sn$ using
SMPN, CWG3M and CWG interactions. Results have been compared with
experimental data.}\label{136sn}

 \end{figure}

It was remarkable that for $^{136,138}Sn$, where the experimental
level  schemes  were  not  known  until recently, the two results
\cite{smp2008}         differ         dramatically         (Figs.
\ref{136sn},\ref{138sn}). The energies of the first three excited
states  2$^+_1$, 4$^+_1$ and 6$^+_1$, for $^{138}Sn$ with the CWG
Hamiltonian appear at about twice of those predicted by SMPN.

 In $^{138}Sn$, multiplet structure with $\nu 2f_{7/2}^6$ is more
clear  in SMPN results for 0$^+$ to 6$^+$ states. Six neutrons or
in other words two neutron holes in $\nu2f_{7/2}$ can  couple  to
$J_{max}=6$.  Thus  for  generating  the  8$^+$  state  a pair of
neutrons must be promoted  to  a  higher  lying  single  particle
orbital.  This  leads  to  an energy gap of about 2.0 MeV, in the
spectrum between the first 6$^+$ and $8^+$ states,  as  shown  in
the SMPN results (Fig.\ref{138sn}). This gap is about 0.8 MeV for
CWG.

While  comparing  \cite{smp2008}  the  structure of the states in
$^{138}Ba$ with six valence protons with that of  the  $^{138}Sn$
with the same number of neutrons, it is found that the protons in
this valence space are more efficient in generating configuration
mixing.   This  is  essentially  due  to  close  spacing  between
$\pi(gds)$  single  particle  orbitals  that  permits   $\pi-\pi$
residual  interaction to scatter easily protons to various single
particle states.

The wave function structure shows \cite{smp2008,smp2010} that the
CWG  interaction  favors  large  configuration  mixing conserving
seniority as far as possible. In contrast, SMPN favors the  purer
structure  of  the  low-lying states and shows characteristics of
$\nu(2f_{7/2})$  multiplets.   Interestingly,   this   structural
difference   has  a  dramatic  effect  in  the  transition  rates
(especially, $B(E2; 2^+ \rightarrow 0^+$) calculated  with  these
two interactions for both $^{136,138}Sn$ \cite{smp2010}.

In  Ref.  \cite{smp2008},  energies,wave  functions,  and  $B(E2,
0^+_{g.s.} \rightarrow 2^+_1)$ values have been compared for  the
yrast  $0^+$  and  $2^+$  states of the three Sn isotopes. It was
found  that  although  the  $E(2^+_1)$  energies  decrease   with
increasing  neutron  number  in  Sn isotopes for N$\geq$84, B(E2)
values corresponding to the SMPN calculations do not increase  as
expected. This feature was already observed experimentally in the
neutron-rich $^{136}Te$ isotope in this mass region. Origin of an
anomalously low $B(E2, 0^+_{g.s.} \rightarrow 2^+_1)$ value found
in  this nucleus was to traced to a reduced neutron pairing above
the N = 82 shell gap. Interestingly,  similar  behavior  is  also
seen  near  the  N  = 20 shell closure for neutron-rich Mg and Ne
isotopes \cite{smp2008}. It seems to be a unique feature observed
in neutron-rich nuclei. In this  domain,  a  comparatively  lower
value  of  $E(2^+_1)$ does not necessarily lead to a larger B(E2)
value indicating collectivity. This type of feature is  predicted
theoretically with the SMPN interaction for N $\geq$ 84.

\subsubsection{The shell closure at N=90 for Sn isotope and its origin}

In  our  next  work  \cite{smp2010},  with  SMPN  interaction, the
prediction for E(2$^+_1$) of  $^{140}Sn$  comes  out  to  be  much
larger  than  the expectation. The comparison of the high 2$^+_1$
energy  of  $^{140}Sn$  with  examples  from  other  neutron-rich
domains  \cite{smp2010}  clearly show that N = 84-88 spectra with
SMPN  manifest  the  effect   of   gradually   filling   up   the
$\nu(2f_{7/2})$  orbital, which finally culminates in a new shell
closure at N = 90.

In  order to understand how the shell closure at N=90 develops as
one increases number of valence  neutrons  from  $^{132}Sn$,  the
effective   single-particle   energies  (ESPE)  for  the  neutron
orbitals  for   the   two   Hamiltonians   have   been   compared
\cite{smp2010}.  To  investigate  the  specific  component of the
interaction which is responsible to generate this new shell  gap,
the  two  body  matrix  elements  of  both CWG and SMPN have been
spin-tensor  decomposed  \cite{smp2010,smp2011}  in  to  central,
antisymmetric spin- orbit (ALS), spin-orbit (LS) and tensor parts
\cite{smp2010}.   For   SMPN,   the  central  and  ALS  part  for
$2f_{7/2}$- $2f_{7/2}$ tbmes account for majority of the downward
shift of the ESPE of $2f_{7/2}$ with increasing  valence  neutron
number  (n).  Variation  in ALS part is primarily responsible for
this observed shell gap at N=90 \cite{smp2010}.

While   analysing   our   results,  we  have  to  understand  the
implication of ALS term. Bare nucleon- nucleon force contains  no
ALS  term.  A  characteristic  feature  common  to many empirical
effective interactions is the strong ALS components in the tbmes.
It  indicates   important   contributions   from   higher   order
renormalisation   or   many   body   effects   to  the  effective
interactions. In empirical SMPN such many -  body  effects  might
have  been  included  in  some  way  through  the modification of
important tbmes.

However,  it  has  been  observed  experimentally  that N = 90 is
suitable for onset of deformation for nuclei above Sn  (like  Xe,
Ba,  etc.).  Therefore,  the energy of the deformed configuration
also comes down. The presence of valence protons above the  inert
core  is  essential for onset of collectivity. The ESPEs for SMPN
which indicate the features of a shell closure for Z = 50 at N  =
90,  have also been analysed to indicate the possibility of onset
of deformation at N = 90 with increasing  Z.  If  the  ESPEs  for
proton  orbitals for N = 90 are plotted, substantial reduction of
the  $1g_{7/2}$  and  $2d_{5/2}$  energy  gap  is  observed  with
increasing   Z,   which   may   favor   onset   of   collectivity
\cite{smp2010}.

\section{The  three body forces and its implication}

It     has     been     discussed    in    our    earlier    work
(\cite{smp2010,smp2011},  and  references   therein)   that   the
realistic  interactions fail to reproduce some shell closures for
neutron-rich nuclei. It has been  shown  \cite{otsuka:1}  that  a
three-body  delta-hole mechanism can explain these shell gaps and
three-body forces are necessary to explain why  the  doubly-magic
$^{24}O$ nucleus is the heaviest oxygen isotope.

Zuker  \cite{zuker}  had  demonstrated earlier that a very simple
three-body monopole term can eliminate  the  limitations  of  the
realistic  two-body  potentials.  According  his prescription, we
have included  corrections  in  the  relevant  tbmes  of  CWG  to
incorporate the three body effects. The correction factor will be
effective for nuclei for which the valence neutron number n= 3 or
more.  It  is  amazing that although CWG does not predict a shell
closure at  $^{140}Sn$,  the  updated  CWG  including  three-body
forces,  named  as CWG3M, also predicts a shell gap for N=90. The
E$(2^+_1$) energy of $^{140}Sn$ predicted by CWG3M (1.889 MeV) is
close    to     that     predicted     by     SMPN     (1.949MeV)
\cite{smp2010,smp2011,smp2012}.  This  also  indicates that three
body effect plays  an  important  role  for  shell  evolution  in
neutron  rich  Sn  isotopes above $^{132}Sn$, as also observed in
$sd$ and $pf$ shells.

\begin{figure}
\vspace{2.51cm}
\hspace{1.5 cm}
\includegraphics[width=0.8\textwidth,angle=0]{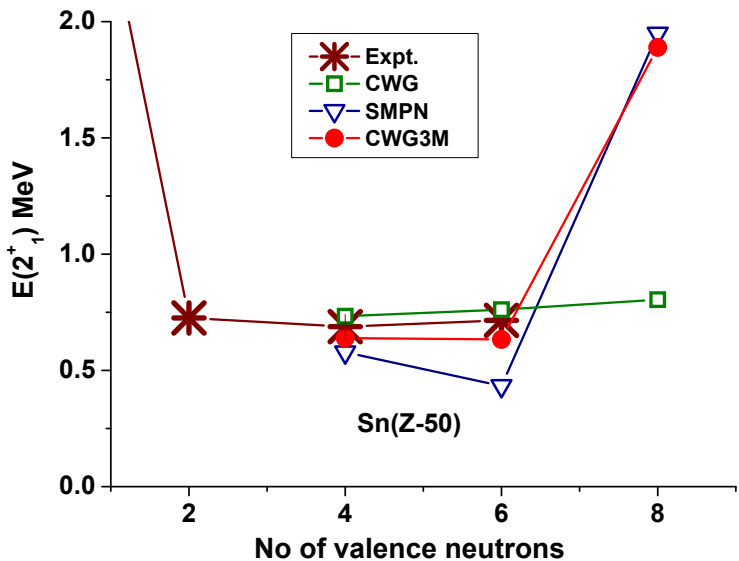}\vspace{-3.51cm}
\caption{  The  calculated $2^+$ energies of $^{134-140}Sn$ using
SMPN, CWG3M and CWG interactions. Results have been compared with
experimental data.}
\label{136sne}
\end{figure}

\section{The new results on $^{136,138}Sn$ and their interpretations}

Very   recently,  experiments  were  carried  out  at  the  RIKEN
Radioactive Isotope Beam  Factory  (RIBF)  \cite{136sn,prl14}  to
study  the neutron rich isotopes of Sn. In one of the experiments
the first $2^+$ excited state in  the  neutron-rich  tin  isotope
$^{136}Sn$  has  been  identified  at  682(13)  keV  by measuring
$\gamma$ -rays in coincidence with the one proton removal channel
from $^{137}Sb$  \cite{136sn}.  In  another  experiment,  delayed
$\gamma$-ray  cascades,  originating  from  the decay of isomeric
states at I=$6^+$, in the very neutron-rich,  semimagic  isotopes
$^{136,138}Sn$   have  been  observed  following  the  projectile
fission of a $^{238}U$ beam at RIBF,  RIKEN.  The  measured  $E_x
(2^+_1)$ value of $^{136}Sn$ was found to be much higher than the
neighboring  N  =  86  isotones,  indicating  a good Z = 50 shell
closure.  The  excitation  energies  stay  almost   constant   in
$^{134}Sn$,  $^{136}Sn$  and  $^{138}Sn$, which suggests that the
seniority - 2 coupling scheme holds beyond N  =  82  up  to  N  =
88\cite{136sn}.  They  found  that the predicted and experimental
$B(E2;  6^+  \rightarrow  4^+$)  values  \cite{prl14}  agree  for
$^{138}Sn$   with  both  empirical  and  realistic  interactions.
However,  for  $^{136}Sn$  the  results  with   their   realistic
interaction,  differ by a factor of $>5$. Three other shell-model
calculations, using realistic and  empirical  interactions,  also
failed  to  reproduce  this  value  for  $^{136}Sn$ by at least a
factor of 2 \cite{prl14}. To  interpret  this  data  better,  the
authors   incorporated   an   empirical   modification   to   the
$2f_{7/2}^2$ matrix elements, probably  the  zero  coupled  ones,
equivalent  to reduced pairing, which generates a seniority-mixed
$4^+_1$  state,  reproducing  all  available  experimental   data
better.  The  authors  proposed  that the wave functions of these
isomeric states \cite{prl14} to be predominantly a fully  aligned
pair of $f_{7/2}$ neutrons.

\begin{figure}
\hspace{-2cm}
\includegraphics[width=.6\textwidth,height=0.85\textheight,angle=-90]{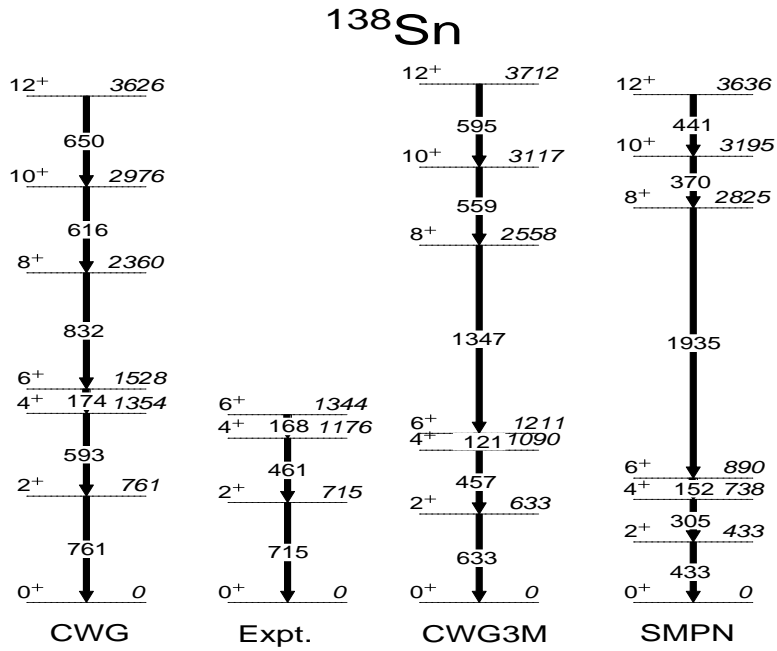}
\caption{  The  calculated excitation spectra of $^{138}Sn$ using
SMPN, CWG3M and CWG interactions. Results have been compared with
experimental data.}
\label{138sn} \end{figure}

\section{Motivation of the present work}
  From the experimental data on odd-even staggering of masses, it
has  been  shown  \cite{pair}  that  variation  of  pairing  as a
function of neutron  number  plays  an  important  role  in  many
distinctive features like occurrence of new shell closures, shell
erosion,  anomalous  reduction  of  the energy of the first $2^+$
state and slower increase of $B(E2; 2^+_1 \rightarrow 0^+_1$)  in
the  neutron-rich  even-even  nuclei  of  different mass regions.
Recently, it has been indicated \cite{holt} that it is  essential
to  include  3N  contributions  to  the pairing interaction for a
quantitative description of nuclear pairing gaps. It  is  usually
found  that  inclusion  of  a  three  body  effect  improves  the
predictive power of realistic interactions. In our earlier  work,
we  have  already  included  three  body effects in realistic CWG
interaction  to  construct  CWG3M.  Our  aim  is  to  study   the
importance  of  three  body  forces  in interpretation of the new
data.

\section{Shell Model Calculations}
In  the  present  work  we  have used the CD-Bonn potential based
realistic CWG \cite{brown} and the empirical SMPN  \cite{smp2004}
along   with   CWG3M   \cite{smp2010}   interaction.   All  these
Hamiltonians have the same set of single-particle energies of the
valence orbitals  but  different  sets  of  two-body  interaction
matrix  elements. The SM calculations \cite{smp2008} were carried
out in the Z50N82 model space using the code OXBASH \cite{oxb:1}.

\begin{table}[ht]
\begin{center}
\begin{tabular}{ccccc} 
\hline
$I_i$ & $I_f$& CWG \cite{brown}&CWG3M \cite{smp2010}&SMPN\cite{smp2004} \\ \hline
\strut \\
\multispan{5} $^{136}Sn$ \hfil\\
2& 0&125&88&90 \\
4&2&86&81&43 \\
6&4&14&45&14 \\
8&6&53&48&46\\
10&8&0.0097&0.0948&0.0954\\
12&10&91&77&69\\\\
6&4(2)&52&16&45 \\
\hline
\strut \\
\multispan{5} $^{138}Sn$\hfil\\
2& 0&184&73&69 \\
4&2&0.3286&45&44 \\
6&4&12&21&20 \\
8&6&1.05&0.20&0.73\\
10&8&138&104&81\\
12&10&60&56&44\\
\hline
\end{tabular}
\caption{Comparison  of  calculated  B(E2) ($e^2fm^4$) values for
major transitions connecting the  positive  parity  yrast  levels
till $12^+$ for $^{136,138}Sn$. The experimental values of $B(E2;
6^+_1   \rightarrow  4^+_1$)  for  $^{136,138}Sn$  is  24(4)  and
$\simeq$ 17 $e^2fm^4$, respectively \cite{prl14}. Effective neutron  charge  is
$e_n=0.64$}

\label{be2}
\end{center}
\end{table}

\section{Results}
We  have  compared  these  new  results with our theoretical ones
calculated using CWG, SMPN and CWG3M  interactions  as  shown  in
Figs.   (\ref{136sne},\ref{136sn},\ref{138sn}).   The  calculated
B(E2) values for all the E2 transitions connecting the yrast band
till 12$^+$ state have also been tabulated in Table \ref{be2}. To
complete the information  -  the  B(E2)  for  $6^+_1  \rightarrow
4^+_2$ for $^{136}Sn$ nucleus has also been tabulated.

\subsection{Energy}
 If  we  keep  in  mind  the usual errors obtained in theoretical
shell model predictions,  for  $^{136}Sn$,  one  can  not  choose
between  these  three  interactions.  However,  the predicted and
experimental  energy  spectra   for   both   $^{136,138}Sn$(Figs.
\ref{136sne},\ref{136sn},\ref{138sn})  show  that  best agreement
has been obtained with the theory in which three body forces have
been included in a realistic interaction (CWG3M).

\subsection{Transition probabilities}
The  measurement  of  transition probabilities and its comparison
with  theory,  is  the  best  way  to  validate   a   theoretical
prediction.   As   discussed  in  Ref.  \cite{smp2008,pair},  the
depresssed $E(2^+_1)$ energies in neutron-rich nuclei arises  due
to  weakening  of  pairing,  unlike  near  stability,  where  low
$E(2^+_1)$ indicate collective deformed state manifested  through
strong B(E2). Therefore, anomalous reduction of the energy of the
first  $2^+_1$  state  and  slower  increase  in the $B(E2; 2^+_1
\rightarrow 0^+_1$) in the neutron-rich  even-even  nuclei  is  a
distinctive  feature  arising out of reduced pairing. It has been
discussed earlier \cite{smp2008}, apart from differences  between
the   $E(2^+_1)$,  the  most  dramatic  difference  lies  in  the
transition  probabilities  of  the  $2^+_1   \rightarrow   0^+_1$
transition  using CWG and SMPN interactions. In the present work,
it is found that the although the  B(E2)  values  for  the  $6^+$
isomers  in  $^{136,138}Sn$ are almost similar with CWG and SMPN,
that from CWG3M is quite different. On the other hand, the B(E2)s
for $2^+_1$ states are similar for both the isotopes are  similar
with CWG3M and SMPN, whereas it is almost double with CWG.

\section{Conclusion}
The  SMPN  and  CWG3M  include the effects of reduced pairing and
three body effects. It has been indicated \cite{136sn,prl14} that
reduced pairing is responsible for the anomalous results in these
nuclei. Our results clearly show that the most sensitive probe to
confirm this claim will be the $B(E2; 2^+_1  \rightarrow  0^+_1$)
values. Thus the importance of pairing will be confirmed from the
transition  probability  measurement  in  these  neutron  –  rich
isotopes of $Sn$.

\section{Acknowledgments}
 Special thanks are owed to B. A. Brown for his help in providing
us the OXBASH (Windows Version) and the NUSHELL@MSU Codes.

\bibliographystyle{pramana}
\bibliography{references}

\begin{thebibliography}{99}
\bibitem{nndc} http://www.nndc.bnl.gov

\bibitem{smp2001}  Sukhendusekhar  Sarkar,  M. Saha Sarkar, Phys.
Rev. C {\bf 64}, 014312 (2001).

 \bibitem{oxb:1}  B. A. Brown, A. Etchegoyen, N. S. Godwin, W. D.
M. Rae, W. A. Richter, W. E.  Ormand,  E.  K.  Warburton,  J.  S.
Winfield,  L.  Zhao  and  C. H. Zimmerman, Oxbash for Windows PC,
MSU-NSCL Report No. {\bf 1289}, (2004); B. A. Brown and W. D.  M.
Rae,NUSHELL@MSU, MSU-NSCL Report,2007.

\bibitem{chou}  W.T.  Chou  and E.K. Warburton, Phys. Rev. C {\bf
45}, 1720 (1992)

 \bibitem{smp2004}  Sukhendusekhar  Sarkar,  M. Saha Sarkar, Eur.
Phys. Jour. A {\bf 21}, 61 (2004).

\bibitem{daess}S.  Sarkar,  M.  Saha Sarkar, Proc. DAE-BRNS Symp.
Nucl.     Phys.     (India)     {\bf     55},     I27     (2010);
(http://www.sympnp.org/proceedings/index.php:  electronic version
only).

\bibitem{136I}   W.   Urban,   M.  Saha  Sarkar,  S.  Sarkar,  T.
Rzaca-Urban, J.L. Durell, A.G. Smith, J.A. Genevey, J.A. Pinston,
G.S. Simpson, I. Ahmad, Euro. Phys. Jour. A {\bf 27}, 257 (2006).

\bibitem{138I}   T.   Rzaca-Urban,  K.  Pagowska,  W.  Urban,  A.
Zlomaniec, J. Genevey, J. A. Pinston,  G.  S.  Simpson,  M.  Saha
Sarkar,  S.  Sarkar,  H. Faust, A. Scherillo, I. Tsekhanovich, R.
Orlandi, J. L. Durell, A. G. Smith, and I. Ahmand, Phys.  Rev.  C
{\bf 75}, 054319 (2007).

\bibitem{138Cs}  T.  Rzaca-Urban,  W.  Urban,  M. Saha Sarkar, S.
Sarkar, J.L. Durell, A.G. Smith, B.J. Varley, and I. Ahmad, Euro.
Phys. Jour. A {\bf 32}, 5 (2007).

\bibitem{brown}  B. A. Brown {\it et al.}, Phys. Rev. C {\bf 71},
044317 (2005).

\bibitem{covello}  A.  Covello,  L.  Coraggio, A. Gargano, and N.
Itaco, J. Phys.Conf. Ser. {\bf 267}, 012019 (2011).

\bibitem{kartamyshev}   M.   P.   Kartamyshev,  T.  Engeland,  M.
Hjorth-Jensen, and E.  Osnes,  Phys.  Rev.  C  {\bf  76},  024313
(2007).

\bibitem{smp2008} S. Sarkar and M. Saha Sarkar, Phys. Rev. C {\bf
78}, 024308 (2008).

\bibitem{smp2010} S. Sarkar and M. Saha Sarkar, Phys. Rev. C {\bf
81}, 064328 (2010).

\bibitem{smp2011}  S.  Sarkar  and  M. Saha Sarkar, J. Phys.Conf.
Ser. {\bf 267}, 012040 (2011).

 \bibitem{otsuka:1}  Takaharu  Otsuka,  Toshio  Suzuki,  Jason D.
Holt, Achim Schwenk and Yoshinori Akaishi, Phys. Rev. Lett.  {\bf
105}, 032501 (2010).

 \bibitem{zuker}  A.  P. Zuker, Phys. Rev. Lett. {\bf 90}, 042502
(2003).

\bibitem{smp2012}  M.  Saha  Sarkar,  S.  Sarkar,  AIP Conference
Proceedings, Volume {\bf 1444}, 117 (2012).

 \bibitem{136sn}  He  Wang  {\it et al.}, Prog. Theor. Exp. Phys.
023D02 (2014).

\bibitem{prl14} G. S. Simpson {\it et al.}, Phys. Rev. Lett. {\bf
113}, 132502 (2014).

\bibitem{pair}  Maitreyee  Saha  Sarkar,  Sukhendusekhar  Sarkar,
Proc. of the Fifth International Conference: Sanibel Island,  USA
4  -10  November  2012, Editors: J H Hamilton, A V Ramayya, World
Scientific, 498 (2013).

\bibitem{holt}  J.D. Holt, J. Menendez and A Schwenk, J. Phys. G:
Nucl. Part. Phys. {\bf 40}, 075105 (2013).

 \end{thebibliography}

\end{document}